\documentclass
[twocolumn,superscriptaddress,showkeys,amsmath,amssymb,apl,nopacs]{revtex4}
\usepackage{graphicx}
\usepackage{dcolumn}
\usepackage{bm}

\usepackage{epsf}
\usepackage{citesort}
\usepackage{epsfig}
\usepackage[latin1]{inputenc}   

\usepackage{amsmath}
\usepackage{amsthm}
\usepackage{amssymb}
\usepackage{epsfig}
\usepackage[bf,footnotesize]{caption}

\begin{document}

\title{Self-compensation in highly n-type InN}
\author{C. Rauch}
\email{christian.rauch@aalto.fi} \affiliation{Department of Applied
Physics, Aalto University, P.O. Box 11100, FI-00076 Aalto, Espoo,
Finland}
\author{F. Tuomisto}
\affiliation{Department of Applied Physics, Aalto University, P.O.
Box 11100, FI-00076 Aalto, Espoo, Finland}
\author{P. D. C. King}
\affiliation{School of Physics and Astronomy,
University of St Andrews, North Haugh, St Andrews KY16 9SS, UK}
\author{T. D. Veal}
\affiliation{Stephenson Institute for
Renewable Energy and Department of Physics, University of Liverpool,
Liverpool, L69 4ZF, UK}
\author{H. Lu}
\affiliation{Department of Electrical and Computer Engineering,
Cornell University, 425 Philips Hall, Ithaca, New York 14853, USA}
\author{W. J. Schaff}
\affiliation{Department of Electrical and Computer Engineering,
Cornell University, 425 Philips Hall, Ithaca, New York 14853, USA}
\date{\today}

\begin{abstract}
Acceptor-type defects in highly n-type InN are probed using positron
annihilation spectroscopy. Results are compared to Hall effect measurements and calculated electron mobilities. Based on this, self-compensation in n-type InN is studied and the microscopic
origin of compensating and scattering centers in irradiated and Si-doped InN
is discussed. We find significant compensation through negatively
charged indium vacancy complexes as well as additional acceptor-type defects with no or small effective open volume, which act as scattering centers in highly n-type InN samples.
\end{abstract}

\maketitle
Indium nitride (InN) is a significantly cation-anion mismatched
semiconductor~\cite{King2008} with promising applications in (opto-)electronics and high-frequency/high-power devices ~\cite{Wu2009}. The material possesses a room-temperature bandgap of
$\sim$0.67~eV (Ref.~\cite{Matsuoka2002}) and a high propensity
toward n-type conductivity. This can be explained by the high
position of the Fermi level stabilization energy ($E_{\text{FS}}$) deep in the
conduction band, at $\sim$1.8~eV above the valence band
maximum~\cite{King2008}. The microscopic origin of the n-type
conductivity has been investigated extensively, and H
impurities ($\text{H}_{\text{i}}^{+},
\text{H}_{\text{N}}^{2+}$)~\cite{Janotti2008} as well as N vacancies
($V_{\text{N}}^{+/3+}$)~\cite{Stampfl2000} are considered the most
dominant contributions in the bulk. At the surface and interface,
contributions from the surface electron
accumulation layer~\cite{Mahboob2004} and charged dislocation
lines~\cite{Piper2006,Bierwagen2011,Miller2011}, respectively, have to be also considered~\cite{King2009}. At low electron
concentrations, the formation energies of donor-type defects are
significantly lower than for acceptors~\cite{Duan2009}. With
increasing Fermi level ($E_{\text{F}}$), however, acceptor incorporation
should become more favorable.
According to density functional theory (DFT) calculations,
$V_{\text{In}}^{3-}$ and $V_{\text{In}}$-complexes are the most
favorable negatively charged centers~\cite{Duan2009a} in n-type InN.
For the highest electron concentrations close to $E_{\text{FS}}$, also
negatively charged single $V_{\text{N}}$~\cite{Walle2010} and N
vacancy clusters~\cite{Duan2008} have been predicted.\\
The main relevance of acceptor-type defects in highly n-type
material is twofold: they (i) form traps for donor-released free
electrons and (ii) act as scattering centers.
Under n-type conditions the
formation energies for multiply charged acceptors (donors) decrease
(increase) rapidly and they tend to become favorable (unfavorable) compared
to the singly charged configuration~\cite{Walle2010}. Compensating
acceptors can play a significant role in reducing the carrier
mobility.\\
The spectroscopic identification of point defects in narrow band gap, highly conducting semiconductors like InN is very challenging. Although, for example, Hall measurements can give indirect evidence of defect charge and concentrations~\cite{Look2002,Jones2007,Hsu2007,Bierwagen2011,Miller2011}, defect identification is mostly not possible. Positron annihilation spectroscopy is a powerful tool for the investigation of vacancies and acceptor-type defects in semiconductors, and is largely not affected by aforementioned properties. Due to reduced Coulomb repulsion, positrons can get trapped at open volume sites in the crystal lattice. This narrows the momentum distribution of annihilating electron-positron pairs which can be measured by recording the Doppler broadened line-shape of the 511 keV annihilation $\gamma$-radiation.\\
In this letter, we use positron annihilation spectroscopy and Hall
effect measurements to determine the dominant compensation
mechanisms in n-type InN.
\begin{table}[b]\label{sampes}
\caption{Free electron concentration of the investigated set of InN samples and fitted positron trapping rates at the dominant vacancy-type positron trap.}
\begin{ruledtabular}
\begin{tabular}{cccc}
ID & Sample & n$_{\text{e}}$ (cm$^{-3}$) & $\kappa_{\text{V}}$ $(s^{-1})$ \\[3pt]
\hline\\
1 & Si-doped & 4.5$\times10^{19}$  & 6.1$\times10^{9}$\\[3pt]
2 & Si-doped & 1.3$\times10^{20}$ & 9.9$\times10^{9}$\\[3pt]
3 & Si-doped & 4.0$\times10^{20}$ & 4.7$\times10^{10}$\\[3pt]
4 & Si-doped & 6.6$\times10^{20}$& 9.4$\times10^{11}$\\[3pt]
5a & He-irr. ($\phi = 8.9 \times 10^{15}cm^{-2}$)& 3$\times10^{20}$&
1.9$\times10^{10}$\\[3pt]
5b & He-irr., RTA & 6$\times10^{19}$& 4.7$\times10^{9}$\\[3pt]
\end{tabular}
\end{ruledtabular}
\label{samples}
\end{table}
Si-doped InN films with a thickness of
$\sim$500nm were grown by plasma-assisted molecular beam epitaxy
(PAMBE) on sapphire substrates with GaN and AlN buffer layers. Free
electron concentrations increase with increasing Si supply from
4.5$\times10^{19}$ cm$^{-3}$ to 6.6$\times10^{20}$ cm$^{-3}$.
Results are compared to those obtained in undoped samples that were
irradiated with 2~MeV $^{4}$He$^{+}$ ions and subsequently treated
by rapid-thermal-annealing (RTA) at temperatures of 425--475$^{\circ}$C~\cite{Reurings2010}. An overview of the investigated set of samples is given
in table~\ref{samples}.\\
Temperature dependent Doppler broadening measurements~\cite{Saarinen1998} have been performed with a variable
energy slow-positron beam. The Doppler broadening of the
$\gamma$-radiation has been recorded by high-purity Ge detectors and analyzed
using the conventional S and W parameters~\cite{Rauch2011c}.
\begin{figure}[b] \centering
\includegraphics[width=0.8\linewidth]{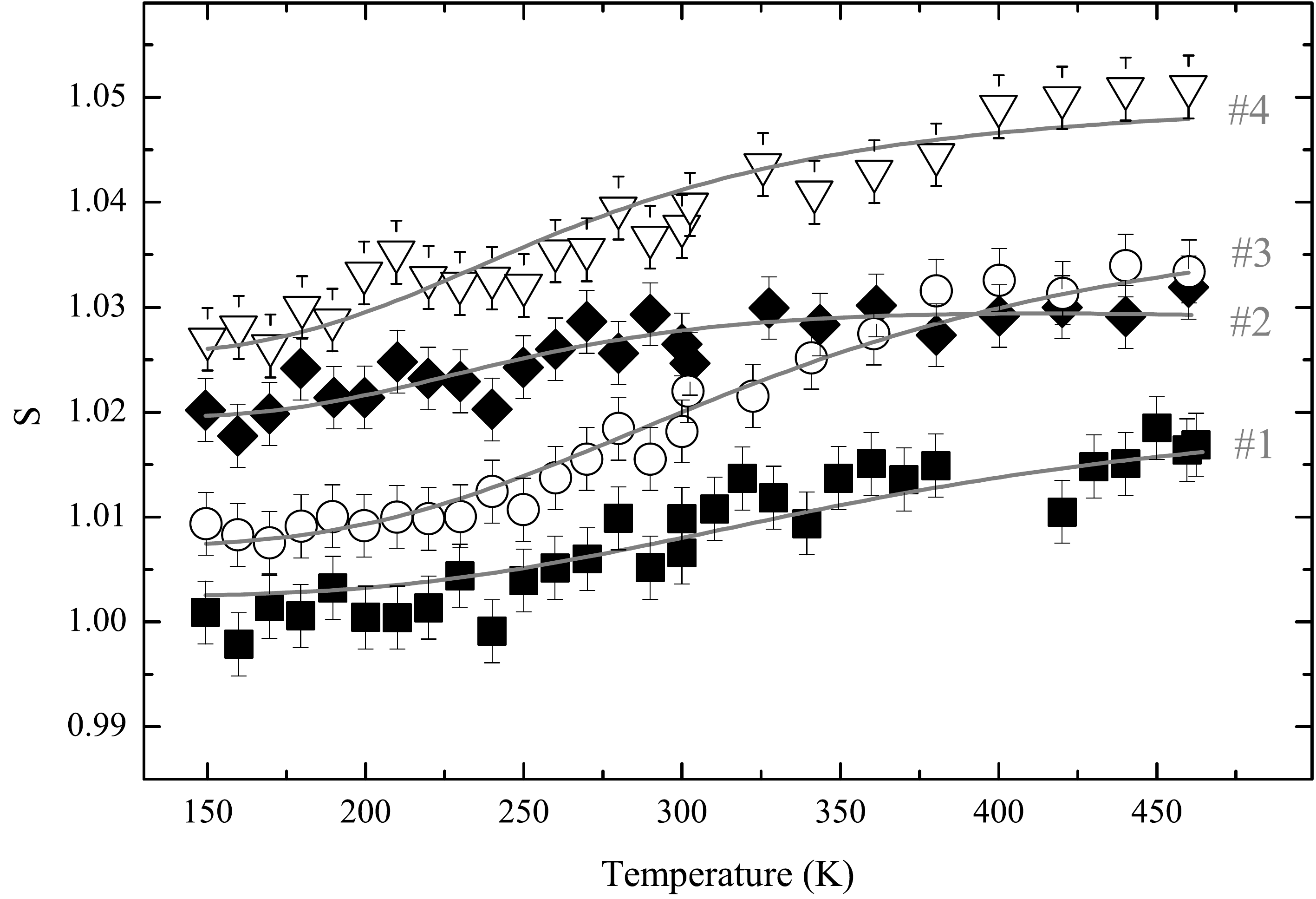}
  \caption{(Color online) Measured temperature dependent S-parameters of Si-
  doped InN samples 1--4. Solid lines are fits assuming competitive trapping between a negatively charged
vacancy defect and one shallow trap.}
  \label{S_Temp}
\end{figure}
The measured relative S-parameters (i.e., normalized to the S-parameter obtained in a reference sample for the InN lattice) in
the Si-doped InN samples for the temperature range of 150--460~K are shown in Fig.~\ref{S_Temp}. All measurements were performed for a mean positron implantation depth of $\bar{x}$=100~nm. A decrease of the S-parameter at low temperatures is visible. This is characteristic for competitive trapping of positrons between deep
vacancy-type traps with high S-parameters and shallow traps with
S-parameters similar to the defect-free lattice. Thermal escape of
positrons from loosely bound defect states, as well as a
T$^{-1/2}$-dependence of the positron trapping coefficient $\mu$ for
the case of charged defects, lead to an overall
temperature dependence of the positron annihilation spectrum~\cite{Puska1990}.\\
The positron de-trapping rate $\delta$ is defined as
\begin{equation}
\delta(T)=\mu(\frac{m_{+}^{*}k_{\text{B}}T}{2\pi\hbar^{2}})^{3/2}\exp(-\frac{E_{\text{b}}}{k_{\text{B}}T}),
\end{equation}
with $m_{+}^{*}$ the effective positron mass, and $E_{\text{b}}$ the
positron binding energy to the respective defect. For the case of
one negatively charged vacancy and one shallow trap with characteristic relative S-parameters of $S_{\text{ST}}=S_{\text{Bulk}}=1$, the measured S-parameter is given as
\begin{equation}\label{PosTempFit}
S(T)=1+\frac{\kappa_{\text{V}}(T)}{\frac{1}{\tau_{\text{Bulk}}}+\frac{\kappa_{\text{ST}}(T)}{1+\tau_{\text{ST}}\delta_{\text{ST}}(T)}+\kappa_{\text{V}}(T)}(S_{V}-1),
\end{equation}
where $\tau_{Bulk}$ and $\tau_{ST}$ are the positron lifetimes for the bulk
and shallow trap, and $\kappa_{V}$ and $\kappa_{ST}$ are the positron trapping rates at vacancy
and shallow trap, respectively.\\
The solid lines in Fig.~\ref{S_Temp} show fits of the spectra using Eq.~\ref{PosTempFit} with $m_{+}^{*}=1$, $\tau_{\text{Bulk}} = \tau_{\text{ST}} \approx 180$~ps
and a relative S-parameter of $S_{V}=1.051$ for the vacancy~\cite{Rauch2011c}. The shallow trap binding energy was
fitted as $E_{b}=90$ meV. Resulting vacancy trapping rates for the Si-doped InN samples are
shown in table~\ref{samples} and range from 6.1$\times10^{9}$ to
9.4$\times10^{11}$~s$^{-1}$ at 300~K. The fitted trapping rates for shallow traps are
in the order of $\gtrsim4\times10^{10}$~s$^{-1}$. The concentrations of
vacancies, $c_{\text{V}}$, and shallow traps, $c_{\text{ST}}$, are
directly proportional to the determined trapping rates: $\kappa=\mu
c$. Assuming a trapping coefficient of
$\mu_{\text{V}}=\mu_{\text{ST}}=3\times10^{15}$~s$^{-1}$ at room temperature, this translates to
concentrations of 1$\times10^{17}$~cm$^{-3} \leq c_{\text{V}} \leq$
2$\times10^{19}$~cm$^{-3}$ and $c_{\text{ST}}\gtrsim1\times10^{18}$~cm$^{-3}$.
\begin{figure}[t] \centering
\includegraphics[width=0.9\linewidth]{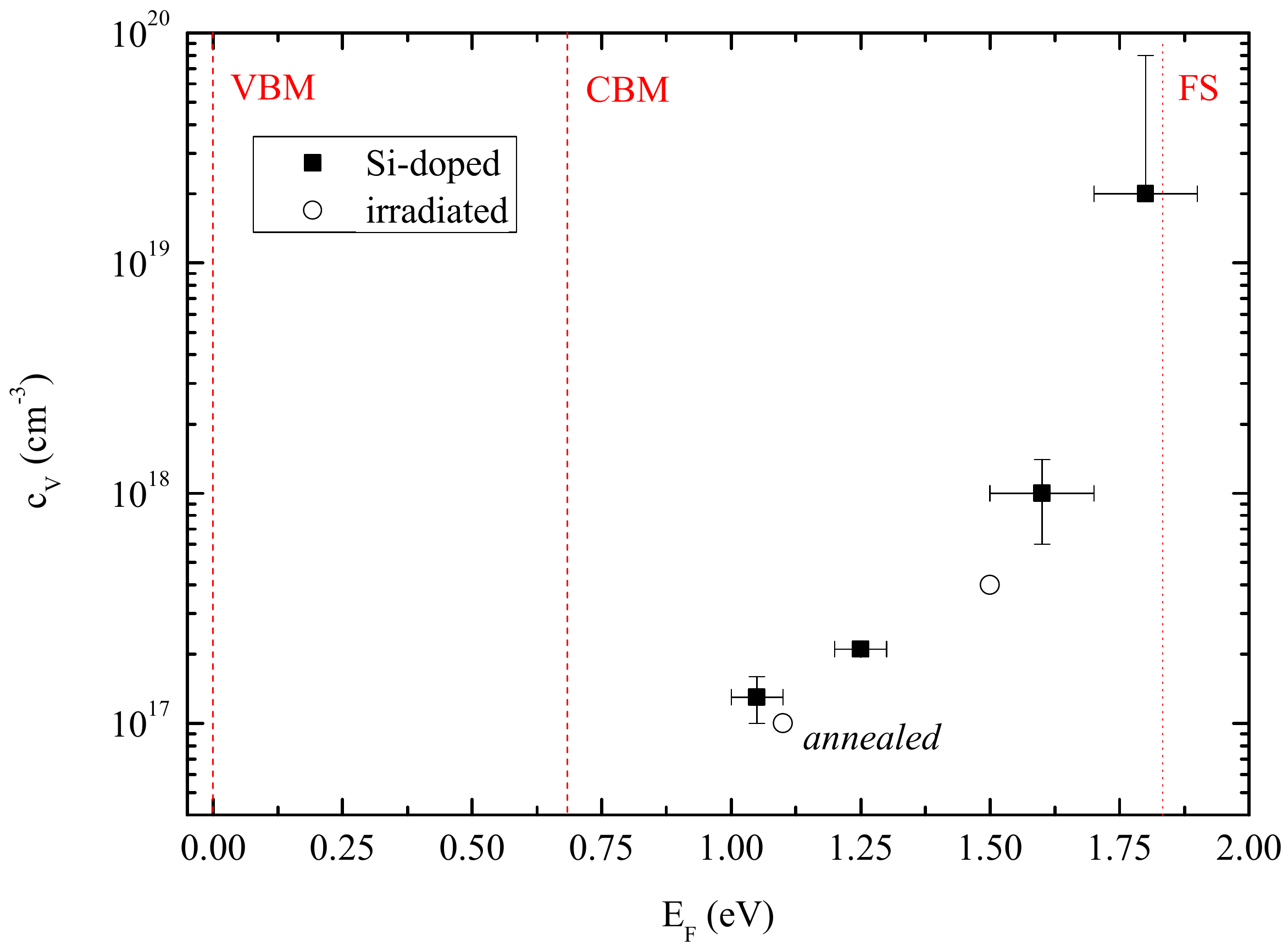}
  \caption{(Color online) Estimated vacancy concentrations of Si-doped
InN samples (full symbols) as a function of the determined bulk Fermi level. Vacancy concentrations for an irradiated sample before and after annealing are given for comparison.}
  \label{VacancyFormation}
\end{figure}
The determined vacancy concentrations are plotted in
Fig.~\ref{VacancyFormation} as a function of the bulk
Fermi level~\cite{King2008} in the respective samples. Previously
estimated defect concentrations of the irradiated
samples~\cite{Reurings2010} are given for comparison. A clear
increase of the vacancy concentrations with increasing Fermi level
is visible, indicating that vacancies are incorporated in the negative charge state. Hence, they should act as efficient scattering
and compensating centers.\\
\begin{figure}[t] \centering
\includegraphics[width=0.9\linewidth]{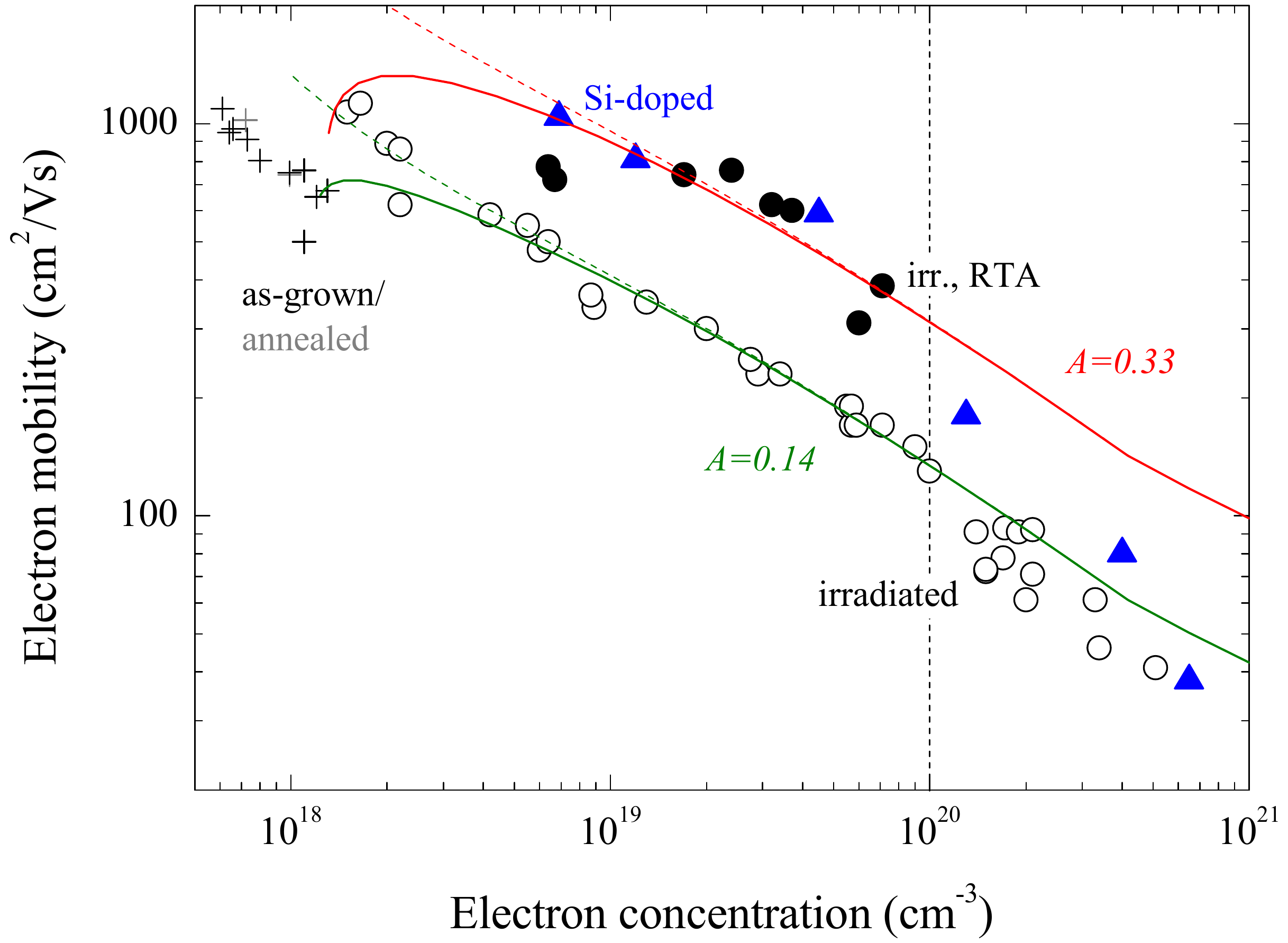}
  \caption{(Color online) Measured electron mobilities of Si doped InN samples (triangles) as a function of the free electron concentration. Mobilities from as-grown (crosses) and irradiated InN samples before (open circles) and after RTA (full circles) are given for comparison (data from Refs.~\protect\cite
{Jones2007,Jones2006b}). Calculated mobilities for two different compensation schemes are shown with (solid lines) and without (dashed lines) the influence of dislocation scattering.}
  \label{Mobility}
\end{figure}
Hall effect measurements of the Si-doped films were performed at room temperature using an
Ecopia HMS-3000 Hall System employing the standard Van der Pauw
configuration. The measured electron mobilities are shown in Fig.~\ref{Mobility}. Mobilities from earlier results of as-grown and irradiated InN layers before and after RTA treatment are given for comparison (data from Refs.~\cite
{Jones2007,Jones2006b}). In the Si-doped films, mobilities range from 40--1050~cm$^{2}$/Vs and are close to the mobilities in the irradiated films after RTA. Mobilities for as-irradiated samples are significantly lower, except for the case of very high electron concentrations. Calculated mobilities are shown in addition, and were determined using a three layer conduction model including near-surface, bulk and interface contributions, and InN reference values according to Ref.~\cite{King2009}. For 500 nm thick films, the influence from the surface is small~\cite{King2009}. Close to the interface, scattering due to charged dislocation lines becomes relevant, and their density was assumed to fall off exponentially with increasing distance from the interface. For more details on the mobility calculations, see Ref.~\cite{King2009}. The bulk mobility ($m$) in n-type InN is dominated by ionized
defect scattering, and is proportional~\cite{Look2011} to
\begin{equation}\label{equ_mobility}
m \sim \frac{n}{Z_{\text{D}}^{2}N_{D}+Z_{\text{A}}^{2}N_{A}}=\frac{Z_{\text{D}}-|Z_{\text{A}}|K}{Z_{\text{D}}^{2}+Z_{\text{A}}^{2}K}=:A,
\end{equation}
where $Z_{\text{D}}$ and $Z_{\text{A}}$ are the donor and acceptor charge,
respectively, and $K=N_{\text{A}}/N_{\text{D}}$ is the compensation ratio. The advantage of using the proportionality constant $A$ for discussing compensation is that the defect charge and compensation ratio do not have to be fixed.\\
At moderate doping levels, the mobility data of the Si-doped samples are fitted well by calculated mobilities using $A = 0.33$. As-irradiated films~\cite{Jones2007} show much lower mobilities and are approximated best with $A=0.14$. After RTA treatment, the mobilities in the irradiated films are close to the values for the Si-doped samples. For low electron concentrations $\lesssim n_{\text{e}}=10^{19}$~cm$^{-3}$, scattering from dislocations is significant. For high doping levels above about $n_{\text{e}}=10^{20}$~cm$^{-3}$ ($E_{\text{F}}=1.1$~eV), the mobility in the Si-doped films starts to deviate strongly from the calculated line for $A = 0.33$ and approaches the irradiated case for the highest doped sample. In the as-irradiated material, a similar decrease in the mobility is observable but less pronounced with a decrease to $A \approx 0.11$. According to Equ.~\ref{equ_mobility}, such a mobility behavior can be explained either by an increase in the compensation ratio or an increase in the acceptor charge.\\
Using Eq.~\ref{equ_mobility}, the densities of dominant donors and acceptors in the InN films can be estimated from the mobility data if their charge is known. In the Si-doped samples, Si$_{\text{In}}^{+}$ can be expected as the dominant donor, with possible contributions from $V_{\text{N}}^{+}$~\cite{Stampfl2000} and $\text{H}_{\text{i}}^{+}$. In the irradiated films, $V_{\text{N}}^{+}$~\cite{Stampfl2000} are likely to be responsible
for the increase in the free electron concentration upon increasing
irradiation doses~\cite{Jones2007}. Therefore, a donor charge of $Z_{\text{D}}=1$ can be estimated. The identity~\cite{Rauch2011c} and
evolution~\cite{Rauch2012c} of dominant vacancy-type acceptors in the InN samples have been probed by positron annihilation spectroscopy and were identified as $V_{\text{In}}$ in the as-irradiated films, and $V_{\text{In}}$-$xV_{\text{N}}$ complexes ($x\approx$~1--3) in the Si-doped samples and irradiated films after RTA treatment.
Additionally, temperature dependent measurements showed a high density of shallow positron traps that can be formed by negatively charged defects with small or no effective open volume. DFT calculations~\cite{Stampfl2000} predict the $V_{\text{In}}$ to be triply negatively charged. For the formation of $V_{\text{In}}$-$xV_{\text{N}}$ complexes, a reduction in the absolute defect charge is expected~\cite{Duan2009a}. The identity of the shallow traps cannot be determined with our experiments. A comparison with DFT calculations~\cite{Duan2008,Walle2010} suggests negatively charged $V_{\text{N}}^{-}$ and its complexes ($xV_{\text{N}}^{-n}$) as well as H$_{\text{N}}^{-}$ as most likely candidates in highly n-type conditions.
Assuming an average charge of $Z_{\text{A}}=-1$ equates to compensation ratios of $K=0.5$ for $A = 0.33$, and $K=0.75$ for $A=0.14$. At a free electron
concentration of for example $n_{\text{e}}=10^{20}$~cm$^{-3}$, this translates to donor and acceptor
concentrations of
$N_{\text{D}}$=$2\times10^{20}$~cm$^{-3}$ and $N_{\text{A}}=1\times10^{20}$~cm$^{-3}$ for $K=0.5$, and $N_{\text{D}}=4\times10^{20}$~cm$^{-3}$ and $N_{\text{A}}=3\times10^{20}$~cm$^{-3}$ for $K=0.75$. Higher acceptor charges correspond to smaller acceptor densities for the same mobility behavior (e.g., $Z_{\text{A}}=-3$, $A=0.14$: $N_{\text{D}}=2.5\times10^{20}$~cm$^{-3}$, $N_{\text{A}}=5\times10^{19}$~cm$^{-3}$).\\
The total acceptor concentrations estimated from Hall effect measurements are about 1--2 orders of magnitude higher than the values determined in temperature dependent Doppler broadening measurements. This may be explained by screening effects in the highly n-type InN samples. Screening reduces long-range Coulomb-related positron capture which could lead to severe underestimation of charged defect concentrations in positron annihilation measurements. This affects especially the estimated densities of shallow traps with no deep positron state. Also for vacancy-type defects the use of a lower positron trapping coefficient may
be appropriate (e.g., $\frac{\mu_{\text{V}^{-}}}{\mu_{\text{V}^{0}}}\approx$3--5,~\cite{Puska1990}), which would lead to higher estimated vacancy densities. An additional factor that could obstruct the exact determination of point defect densities in positron annihilation measurements of n-type InN is the high dislocation densities in the material~\cite{Rauch2012c,King2009}. These might lead to a constant background
trapping of positrons and in turn to an underestimation of the effective
trapping rates of the remaining centers in Doppler
broadening spectroscopy. The exceptionally low positron diffusion
lengths up to high temperatures in the InN samples~\cite{Rauch2012c} speak in favor of such a scenario. A high compensation ratio in the Si-doped InN layers is additionally supported by the observation of strong Urbach tails in optical absorption spectra from these samples~\cite{King2008}.\\
The exponential increase of the concentration of $V_{\text{In}}$ and its complexes, as observed in positron annihilation measurements (Fig.~\ref{VacancyFormation}), coincides with the onset of the strong deviation of the measured electron mobilities in the Si-doped samples from the calculated mobilities at a Fermi energy of $\sim$1.1~eV (Fig.~\ref{Mobility}). This suggests that scattering from acceptor-type defects starts to contribute strongly to the mobility behavior at elevated free electron concentrations. The observed high positron
trapping rates to shallow traps indicate that
$V_{\text{In}}$-related defects do not account alone for the whole
compensation, but additional negatively charged acceptors are
present in the investigated samples in high concentrations. These might be formed by singly negatively charged $V_{\text{N}}$ and multiply negatively charged $V_{\text{N}}$ clusters which become increasingly favorable at high Fermi levels~\cite{Duan2008}. Indirect evidence of the presence of $V_{\text{N}}$ clusters was found by the observation of $V_{\text{In}}$-$xV_{\text{N}}$
complexes ($x\approx$~1--3) in positron measurements of the Si-doped InN samples~\cite{Rauch2011c}. The lower charge of the compensating $V_{\text{In}}$-$V_{\text{N}}$ complexes in Si-doped samples and RTA treated irradiated layers, compared to the triply charged $V_{\text{In}}$ in the as-irradiated samples, might contribute to the observed mobility drop after annealing.\\
The experimentally observed high acceptor densities in highly n-type InN are in sharp contrast to theoretical values based on formation energies from DFT calculations~\cite{Walle2010,Duan2008,Duan2009,Duan2009a}. This suggests that thermal equilibrium considerations might not be appropriate for estimating point defect concentrations in n-type InN, a material which is commonly grown at low temperatures. The formation of point defects during growth of InN is likely to be dominated by other mechanisms~\cite{Rauch2012c}.\\
In conclusion, positron annihilation and Hall effect
measurements have been used to investigate compensation in n-type InN. The densities of acceptor-type defects in Si-doped and
irradiated n-type InN with free carrier concentrations up to $n_{\text{e}}=
6.6\times10^{20}$~cm$^{-3}$ were estimated from temperature-dependent Doppler broadening spectra and compared to Hall mobility data. Significant compensation of n-type InN is found at high Fermi levels, which is attributed to the presence of negatively charged indium vacancy complexes and additional acceptor-type defects with small or no effective open volume. Our results indicate a significant contribution from acceptor-type defects to the mobility behavior in highly n-type InN, while scattering from ionized donors and charged dislocations is dominant at moderate electron concentrations.
\begin{acknowledgements}
This work has been supported by the European Commission under the
7th Framework Program through the Marie Curie Initial Training
Network RAINBOW, Contract No. PITN-Ga-2008-213238, the Engineering and Physical Sciences Research Council, UK under grant no. EP/G004447/1, and the Academy of Finland.
\end{acknowledgements}

\end{document}